\documentclass[article,superscriptaddress,aps,twocolumn,pra,longbibliography]{revtex4-2}
\usepackage{dcolumn}
\usepackage{bm}
\usepackage{epsfig}
\usepackage{graphicx}
\usepackage{amssymb,amsmath,amsbsy,amsgen,amsfonts,amsthm,mathtools}    
\usepackage{mathrsfs}
\usepackage{latexsym}
\usepackage{array}
\usepackage{color}
\usepackage{amstext}
\allowdisplaybreaks[1]
\usepackage{txfonts}
\usepackage{bbold}
\usepackage{pifont}
\usepackage{verbatim}
\usepackage{hyperref}
\usepackage[normalem]{ulem}

\DeclareSymbolFont{symbols}{OMS}{cmsy}{m}{n}

\begin{document}
\title{Distributed quantum phase sensing for arbitrary positive and negative weights}

\author{Changhun Oh}
\email{changhun@uchicago.edu}
\affiliation{Pritzker School of Molecular Engineering, The University of Chicago, Chicago, Illinois 60637, USA}

\author{Liang Jiang}
\affiliation{Pritzker School of Molecular Engineering, The University of Chicago, Chicago, Illinois 60637, USA}

\author{Changhyoup Lee}
\email{changhyoup.lee@gmail.com}
\affiliation{Quantum Universe Center, Korea Institute for Advanced Study, Seoul 02455, Korea}
\affiliation{Korea Research Institute of Standards and Science, Daejeon 34113, Korea}
\date{\today}

\begin{abstract}
Estimation of a global parameter defined as a weighted linear combination of unknown multiple parameters can be enhanced by using quantum resources. Advantageous quantum strategies may vary depending on the weight distribution, requiring the study of an optimal scheme achieving a maximal quantum advantage for a given sensing scenario. 
In this work, we propose a Heisenberg-limited distributed quantum phase sensing scheme using Gaussian states for an arbitrary distribution of the weights with positive and negative signs.
The proposed scheme exploits entanglement of Gaussian states only among the modes assigned with equal signs of the weights, but separates the modes with opposite weight signs. We show that the estimation precision of the scheme exhibits the Heisenberg scaling in the mean photon number and it can be achieved by injecting two single-mode squeezed states into the respective linear beam-splitter networks and performing homodyne detection on them in the absence of loss.
Interestingly, the proposed scheme is proven to be optimal for Gaussian probe states with zero displacement.
We also provide an intuitive understanding of our results by focusing on the two-mode case, in comparison with the cases using non-Gaussian probe states. We expect this work to motivate further studies on quantum-enhanced distributed sensing schemes considering various types of physical parameters with an arbitrary weight distribution. 
\end{abstract}


\maketitle

\section{Introduction}

Quantum sensing enables more precise estimation of unknown parameters than what is possible with classical resources~\cite{ Giovannetti2011}. A quantum enhancement is only achieved using an appropriate combination of a probe and a measurement, and can thus be maximized by the use of the optimal quantum resource~\cite{Toth2014, Demkowicz-Dobrzanski2015}. Therefore, it is of utmost importance to identify the optimal quantum probe state and measurement setting in order to achieve the ultimate quantum limit leading to the maximum quantum enhancement~\cite{paris2009quantum}. Since the pioneering work of Caves~\cite{Caves1981}, a number of quantum sensing and metrological techniques have been developed and experimentally demonstrated in various sensing scenarios in diverse physical systems~\cite{Degen2017, Braun2018, Pirandola2018, Pezze2018, Lee2021}.

While most studies have focused on an estimation of a single unknown parameter, recent researches on multiple parameter estimation started to attract intensive interest from the quantum sensing community for unrevealed fundamental questions and practical perspectives~\cite{zhuang2018distributed, liu2019quantum, gessner2020multiparameter, qian2021optimal, kwon2022quantum}. 
The main question is if quantum correlation of a probe state is advantageous when multiple parameters are estimated simultaneously as compared to estimating them individually~\cite{proctor2018multiparameter}. Upon the derivation of quantum Cram{\'e}r-Rao matrix inequality for the covariance of estimates of multiple parameters~\cite{liu2019quantum}, the question has been answered 
in particular sensing scenarios by using various quantum states such as a coherent superposition of $N$ photons~\cite{Humphreys2013}, Gaussian states~\cite{Gagatsos2016} or particle-mode-entangled states~\cite{Gessner2018}. A more tricky scenario has also been discussed, e.g., estimating multiple phases governed by non-commuting generators~\cite{Baumgratz2016, Hou2020}.

The use of entanglement does not always promise a quantum enhancement in simultaneous estimation~\cite{Knott2016, proctor2018multiparameter}, but the role of entanglement becomes significant in estimating a global parameter composed of multiple parameters that are encoded across multiple modes or locations~\cite{proctor2018multiparameter, ge2018distributed, Gessner2018, oh2020optimal, gessner2020multiparameter, kwon2022quantum, Xia2020, Guo2020, qian2021optimal, Humphreys2013, Liu2021}, called distributed sensing. 
Distributed quantum sensing has various applications such as global clock synchronization \cite{komar2014quantum}, phase imaging \cite{Humphreys2013, albarelli2020perspective}, and detection of radio-frequency signals \cite{Xia2020}.
The most common type of a global parameter having been of interest in distributed sensing is a linear combination of multiple parameters with weights~\cite{Rubio2020, Gross2021, Liu2021, proctor2018multiparameter, oh2020optimal, ge2018distributed}. 
The weights determine not only the optimal allocation of modal energies over the modes, but also the type of an optimal state. The particular weight distributions have been considered in several theoretical studies~\cite{Gatto2019,triggiani2021heisenberg, oh2020optimal, Guo2020}. For equal positive weights, a quantum enhancement has been experimentally demonstrated in a scheme using a squeezed vacuum state being injected into a beam splitter array for an estimation of the average phase~\cite{Guo2020} and the average displacement parameter~\cite{Xia2020}. For unequal weights, entangled photons have been used to achieve a reduced noise below the shot noise limit with post-selection~\cite{Liu2021} and without post-selection~\cite{zhao2020field}. Furthermore, a linear combination of displacements with unequal weights have been measured using a squeezed light~\cite{Xia2020}.
In general, entanglement is known to be advantageous for a global parameter estimation~\cite{proctor2018multiparameter} and shown to achieve the Heisenberg scaling in distributed phase sensing for arbitrary weights \cite{proctor2018multiparameter,ge2018distributed}.
However, the optimal scheme to achieve the Heisenberg scaling needs to be identified particularly for continuous variable systems such as Gaussian states, although a Gaussian scheme with entanglement attaining the Heisenberg scaling has been studied very recently~\cite{Malitesta2021}.

In this work, we consider distributed quantum sensing to estimate a linear combination of phases with arbitrary positive and negative weights.
We propose a Heisenberg-limited scheme using Gaussian states that takes into account both arbitrary magnitudes of the weights and their signs and find an achievable estimation error.
Especially we prove that our scheme is optimal when we restrict input states to Gaussian states with zero displacement.
It is interesting that the optimal zero-displacement Gaussian scheme decomposes the modes into two groups according to the sign of associated weights, i.e., no entanglement between the two groups, but only within the individual groups. To further understand the role of entanglement, we elaborate on the two-mode case, where the two phases are linearly combined with arbitrary signs.
Through the numerical optimization performed for the two-mode case, we show that the optimal scheme employs neither entanglement nor displacement when the weights have the opposite signs. 
This result is more general than the result in Ref.~\cite{lang2014optimal}, where the product of two single-mode squeezed states is shown to be the optimal input when only the phase difference is unknown in a two-mode interferometer.
We also discuss the origin of the above behavior through the comparison with the cases using (non-Gaussian) entangled photons. 


\section{Results}
\subsection{Multiparameter estimation theory}
Let us consider an estimation of $M$ parameters $\bm{\phi}=(\phi_1,\phi_2,...,\phi_M)^\text{T}$ using measurement outcomes~$\bm{x}$, following a conditional probability~$p(\bm{x}\vert\bm{\phi})$.
The multiparameter Cram\'{e}r-Rao inequality imposes a lower-bound of the~$M\times M$ estimation error matrix~$\Sigma_{ij}=\langle (\hat{\phi}_i-\phi_i)(\hat{\phi}_j-\phi_j) \rangle$ of any unbiased estimator~$\hat{\phi}_i$ by the Fisher information matrix,~$\bm{F}(\bm{\phi})$, i.e.,~$\bm{\Sigma} \geq \bm{F}^{-1},$ where~
$\bm{F}_{ij}(\bm{\phi})=\sum_{\bm{x}} \frac{1}{p(\bm{x}\vert\bm{\phi})}\frac{\partial p(\bm{x}\vert \bm{\phi})}{\partial\phi_i}\frac{\partial p(\bm{x}\vert\bm{\phi})}{\partial\phi_j}$~\cite{helstrom1976quantum}.
In quantum estimation theory, the quantum Cram\'{e}r-Rao matrix inequality gives a lower bound for the error of any unbiased estimator, i.e., $\bm{\Sigma} \geq \bm{F}^{-1} \geq \bm{H}^{-1}$,
where~$H_{ij}=\text{Tr}[\hat{\rho}_{\bm{\phi}}\{\hat{L}_i,\hat{L}_j\}]/2$ is the quantum Fisher information matrix (QFIM), with~$\hat{L}_i$ being a symmetric logarithmic derivative operator associated with~$i$th parameter~$\phi_i$~\cite{braunstein1994statistical}.
Here, $\{\hat{A},\hat{B}\}\equiv\hat{A}\hat{B}+\hat{B}\hat{A}$.
Especially when a linear combination of~$\phi_i$'s is of particular interest, i.e.,~$\phi^*\equiv\boldsymbol{w}^\text{T}\bm{\phi}=\sum_{i=1}^M w_i\phi_i$ with an arbitrary weight vector~$\boldsymbol{w}$, the estimation error is lower-bounded as~\cite{paris2009quantum}
\begin{align}\label{err}
\Delta^2\phi^*\equiv \langle (\hat{\phi}^*-\phi^*)^2\rangle\geq \boldsymbol{w}^\text{T}\bm{F}^{-1}\boldsymbol{w}\geq \boldsymbol{w}^\text{T}\bm{H}^{-1}\boldsymbol{w}, 
\end{align}
where the bound is called the quantum Cram{\'e}r-Rao bound (QCRB). Here, if the matrices are singular, $\bm{F}^{-1}$ and~$\bm{H}^{-1}$ are understood as the inverse on their support.
Throughout this paper, we assume the normalization~$\|\boldsymbol{w}\|_1\equiv \sum_{i=1}^M\vert w_i\vert=1$ for simplicity. 

\subsection{Distributed Gaussian phase sensing}
In this work, we focus on using Gaussian states as a probe to encode multiple parameters $\bm{\phi}$.
Gaussian states are defined as states whose Wigner function follows a Gaussian distribution.
Thus, a Gaussian state $\hat{\rho}$ is fully characterized by its first moment vector~$d_i=\text{Tr}[\hat{\rho}\hat{Q}_i]$ and covariance matrix~$\Gamma_{ij}=\text{Tr}[\hat{\rho}\{\hat{Q}_i-d_i,\hat{Q}_j-d_j\}/2]$. 
Here, a quadrature operator vector of a~$M$-mode continuous variable quantum system is defined as~$\hat{\bm{Q}}=(\hat{x}_1, \hat{p}_1,...,\hat{x}_M,\hat{p}_M)^\text{T}$, satisfying the canonical commutation relation, $[\hat{Q}_j,\hat{Q}_k]=i(\bm{\Omega}_{2M})_{jk}$, where $\bm{\Omega}_{2M}=
\mathbb{1}_M\otimes
\scriptsize{\begin{pmatrix}
0 & 1 \\
-1 & 0
\end{pmatrix}}$ and
$\mathbb{1}_{M}$ is the $M \times M$ identity matrix.

\begin{figure}[b]
\centering
\includegraphics[width=\linewidth]{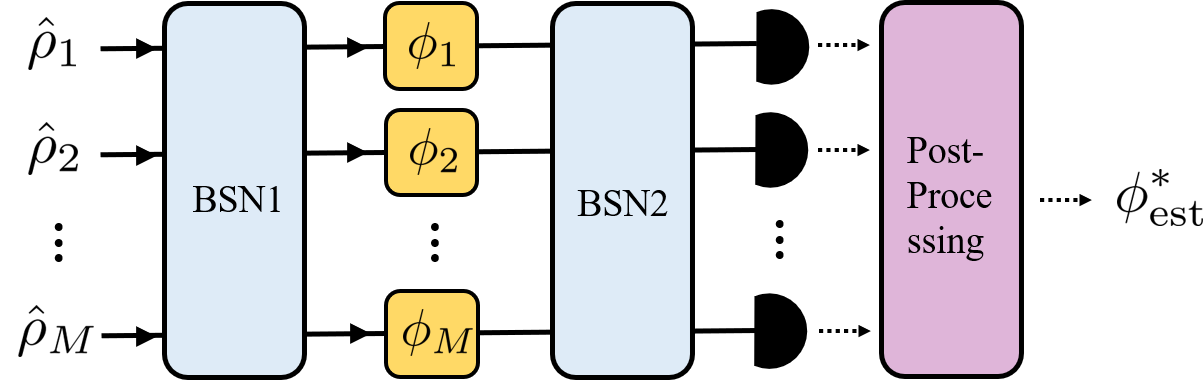}
\caption{Schematic of distributed sensing under investigation. 
A multi-mode probe state $\hat{\rho}_\text{probe}$ generated from the first beam splitter network (BSN) for a given product state input $\otimes_{i=1}^{M}\hat{\rho}_{i}$ undergoes the individual phase shifts on each mode. The parameter-imprinted state $\hat{\rho}_{\bm{\phi}}$ is fed into the second BSN (if necessary), followed by measurement. The measurement outcomes are used in post-processing to estimate the parameter~$\phi^*=\sum_{i=1}^{M}w_i \phi_i$ with the weight vector~$\boldsymbol{w}$.
}
\label{setting}
\end{figure}

We consider a setting shown in Fig.~\ref{setting}, through which any Gaussian probe state~$\hat{\rho}_\text{probe}$ can be prepared by applying a beam splitter network (BSN) to a product Gaussian state input~$\otimes_{i=1}^{M}\hat{\rho}_i$~\cite{reck1994experimental, weedbrook2012gaussian}. Multiple phases are then encoded on the probe state via a unitary operation~$\hat{U}_{\bm{\phi}}=\otimes_{j=1}^{M}e^{-i \phi_j \hat{N}_j}$. 
The output state~$\hat{\rho}_{\bm{\phi}}=\hat{U}_{\bm{\phi}}\hat{\rho}_\text{probe}\hat{U}_{\bm{\phi}}^\dagger$~is finally measured after the second BSN that is inserted to realize a non-local measurement if necessary. 
Here, a strong reference beam is implicitly assumed to define the phases, accessible in each mode for phase-sensitive measurement~\cite{jarzyna2012quantum}. In many cases of quantum sensing, the energy constraint is imposed to the modes that pass through objects whose features are estimated. We thus impose the energy constraint to $M$ modes, i.e., the energy for the reference beam is excluded in accounting of the resource cost.

For the Gaussian probe state  $\hat{\rho}_\text{probe}$ characterized by the covariance matrix $\bm{\Gamma}$ and first moment vector $\bm{d}$, the QFIM in Eq.~\eqref{err} can be written as~\cite{banchi2015quantum, serafini2017quantum, nichols2018multiparameter, oh2019optimal, liu2019quantum, sidhu2020geometric}
\begin{align}\label{qfim}
H_{ij}=&2\text{Tr}[\bm{\Gamma}^{(i,j)}\bm{\Gamma}^{(j,i)}]-\delta_{ij}+(\bm{\Omega}_2 \bm{d}^{(i)})^\text{T}[\bm{\Gamma}^{-1}]^{(i,j)}(\bm{\Omega}_2 \bm{d}^{(j)}),
\end{align}
where~$\bm{A}^{(i,j)}$ denotes the~$2\times 2$ submatrix in the $i$th row and~$j$th column of the $M\times M$ block matrix~$\bm{A}$, and similar for the vector~$\bm{d}^{(i)}$.
The derivation of the QFIM of Eq.~\eqref{qfim} can be found in Ref.~\cite{oh2020optimal}.
We note that considering pure probe states is sufficient to find an optimal state maximizing the QFIM because of its convexity~\cite{ge2018distributed}, while the analytical form of the QFIM for general Gaussian states including mixed states can also be found~\cite{banchi2015quantum, serafini2017quantum, nichols2018multiparameter, oh2019optimal, liu2019quantum, sidhu2020geometric}.
Since the generators of parameters, \{$\hat{N}_i\}_{i=1}^M$, commute, the QCRB in Eq.~\eqref{err} can be saturated~\cite{pezze2017optimal}.
Throughout this work, we assume zero displacement, i.e., $\bm{d}=0$, for which all the off-diagonal elements of QFIM are non-negative. The latter feature is important to analyze the results of this work, as discussed later.

\subsection{Distributed Gaussian phase sensing for arbitrary weights without entanglement}
\textit{Standard quantum limit -- }
The standard quantum limit (SQL) in distributed phase sensing for arbitrary weights is defined by the use of a product coherent probe state, for which the QCRB is written as~\cite{oh2020optimal}
\begin{align}
\Delta^2\phi^*\geq \sum_{i=1}^M \frac{w_i^2}{\bar{N}_i}=\frac{1}{4\bar{N}}.
\label{eq:SQL}
\end{align}
The best strategy for a given total average photon number~$\bar{N}$ is to distribute the average photon number $\bar{N}$ over the modes according to the weight magnitudes~$\vert w_i\vert$, i.e.,~$\bar{N}_i=\vert w_i\vert \bar{N}$, independent of the weight signs. 

\textit{Optimal separable Gaussian scheme -- }
More useful product Gaussian states leading to a smaller error than the SQL of Eq.~\eqref{eq:SQL} can be found and the best strategy under the photon number constraint~$\bar{N}$ 
is to prepare the probe state in a product of single-mode squeezed vacuum states with~$\bar{N}_i^2(\bar{N}_i+1)^2/(2\bar{N}_i+1)\propto w_i^2$ and encode phases without implementing a BSN. 
In this case, the lower-bound of the estimation error becomes
\begin{align}\label{OPGSerror}
\Delta^2\phi^*\geq \sum_{i=1}^M\frac{w_i^2}{8\bar{N}_i(\bar{N}_i+1)},
\end{align}
where individual modes scale with $\bar{N}_i^2$, i.e., Heisenberg scaling is achieved. One can also show that the QCRB bound of Eq.~\eqref{OPGSerror} can be achieved by performing homodyne detection on each mode without the second BSN~\cite{olivares2009bayesian}.

\begin{figure}[b]
\centering
\includegraphics[width=\linewidth]{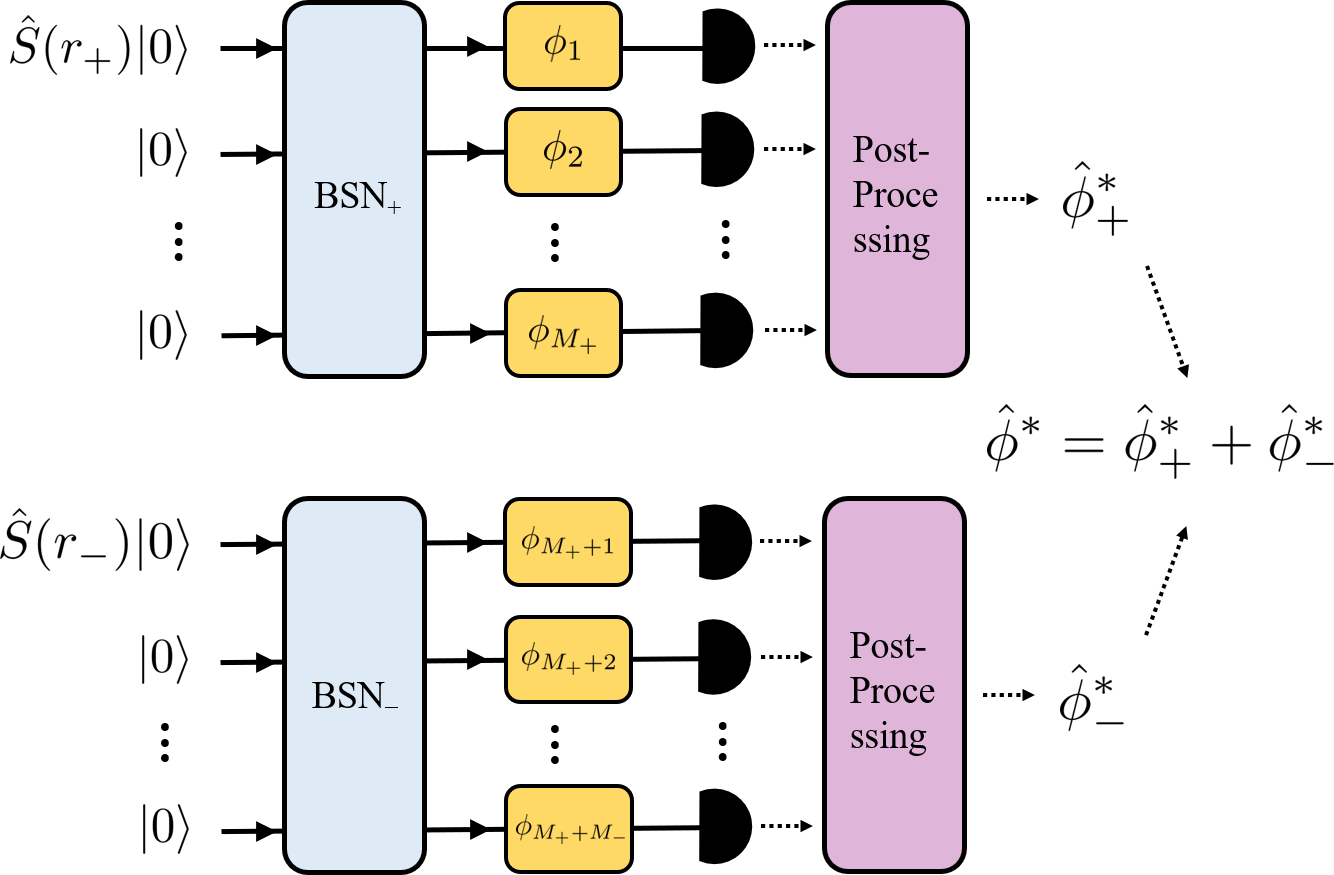}
\caption{Optimal scheme to estimate an arbitrary linear combination of phases $\phi^*$.
Here, $\phi^*\equiv \sum_{i=1}^M w_i\phi_i$ with $w_i>0$ for $1\leq i\leq M_+$ and $w_i<0$ for $M_+< i\leq M_++M_-=M$, i.e., $\phi^*_+=\sum_{i=1}^{M_+}w_i\phi_i$ and $\phi^*_-=\sum_{i=M_++1}^{M_++M_-}w_i\phi_i$.
We constitute two independent BSNs ($\text{BSN}_\pm$) and squeezed vacuum states to estimate $\phi_+^*$ and $\phi_-^*$ separately and estimate $\phi^*$ by their sum, $\phi^*=\phi_+^*+\phi_-^*$.
Finally, homodyne detection on each site is performed to achieve the optimal precision.
Thus, the second BSN in Fig. \ref{setting} is not necessary.
}
\label{proposed_setup}
\end{figure}

\subsection{Optimal entangled scheme for Gaussian states with zero displacement}
We now consider a more general case where an {\it entangled} zero-mean Gaussian state is employed.
Assuming $w_i\neq 0$ for all $i$'s without loss of generality, let us decompose the weight vector $\boldsymbol{w}$ into a positive part $\boldsymbol{w}_+$ and a negative part $\boldsymbol{w}_-$, so that $\boldsymbol{w}=\boldsymbol{w}_++\boldsymbol{w}_-$, i.e., all non-zero elements of $\boldsymbol{w}_+$ ($\boldsymbol{w}_-$) are positive (negative). Let $M_\pm$ be the number of modes corresponding to the weights $\boldsymbol{w}_\pm$. Grouping the modes corresponding to $\boldsymbol{w}_\pm$ (i.e., two groups), we propose a scheme that treats the two groups independently, i.e., first estimating $\phi_{\pm}^{*}=\boldsymbol{w}_{\pm}^{\text{T}}\boldsymbol{\phi}$ individually and finally calculating $\phi^{*}=\phi_+^{*}+\phi_-^{*}$, which is illustrated in Fig. \ref{proposed_setup}.
The estimation error of the particularly proposed scheme for $\phi^*$ is given simply by the sum of individual optimal estimation errors of $\phi_+^*$ and $\phi_-^*$ [see Eq.~\eqref{eq:QCRB_individual}],
i.e.,
\begin{align}\label{eq:err_sum}
\Delta^2\phi^*= \frac{\|\boldsymbol{w}_+\|_1^2}{8\bar{N}_+(\bar{N}_++1)}+\frac{\|\boldsymbol{w}_-\|_1^2}{8\bar{N}_-(\bar{N}_-+1)},
\end{align}
where $\bar{N}_\pm$ represent the average photon numbers of the optimal entangled states used for probing in the respective groups.
Interestingly, the error of Eq.~\eqref{eq:err_sum} is the same as the QCRB that is obtainable by a globally optimal scheme, whose proof is provided in Sec.~\ref{sec:optimal}. This means that the proposed scheme is optimal over all schemes using Gaussian states with zero displacement.
One can easily show that if we choose $\bar{N}_\pm=\bar{N}/2$, the QCRB of Eq.~\eqref{eq:err_sum} follows the Heisenberg scaling, i.e., it scales as $1/2\bar{N}^2$.
A further optimization can be made by optimally allocating $\bar{N}_\pm$ under the constraint $\bar{N}=\bar{N}_++\bar{N}_-$.
The optimal allocation can be found to be $\bar{N}_\pm^2(\bar{N}_\pm+1)^2/(2\bar{N}_\pm+1)\propto \|\boldsymbol{w}_\pm\|_1^2$ by using the Lagrange multiplier method.
While further analytical simplification of the QCRB with the optimal photon number allocation $\bar{N}_\pm$ is cumbersome, we show in Appendix \ref{appendix:upper} that the QCRB minimized by optimal $\bar{N}_\pm$ is upper-bounded as
\begin{align}\label{eq:min}
    \frac{\|\boldsymbol{w}_+\|_1^2}{8\bar{N}_+(\bar{N}_++1)}+\frac{\|\boldsymbol{w}_-\|_1^2}{8\bar{N}_-(\bar{N}_-+1)}< \frac{1}{4\bar{N}^2}.
\end{align}
Thus, it is evident that the QCRB follows the Heisenberg scaling in $\bar{N}$.
It also becomes clear that the QCRB of Eq.~\eqref{eq:err_sum} is always smaller than or at least equal to the one in Eq.~\eqref{OPGSerror} by showing that the optimal precision for each group is given by $\|\boldsymbol{w}_\pm\|_1^2/[8\bar{N}_\pm(\bar{N}_\pm+1)]$.


\textit{Optimality for the individual group with equal weight signs -- }
The derivation of the optimal sensitivity for the individual groups is given as follows.
Denoting $\boldsymbol{w}=\boldsymbol{w}_{\pm}$ and $M=M_\pm$ for convenience of the derivation, we further develop the QCRB of Eq.~\eqref{err} as 
\begin{align}
\Delta^2 \phi^*
&\geq\boldsymbol{w}^\text{T}\bm{H}^{-1}\boldsymbol{w}\nonumber\\
&\geq \frac{(\boldsymbol{w}\cdot \boldsymbol{v})^2}{\boldsymbol{v}^\text{T}\bm{H}\boldsymbol{v}}
=\frac{(\boldsymbol{w}\cdot \boldsymbol{v})^2}{4(\Delta^2 \hat{G}')_\psi} \nonumber\\
&\geq \frac{(\boldsymbol{w}\cdot\boldsymbol{v})^2}{4 \max_\psi (\Delta^2 \hat{G}')_\psi} \nonumber \\ 
&=\frac{\|\boldsymbol{w}\|_1^2}{8\bar{N}(\bar{N}+1)},\label{eq:QCRB_individual}
\end{align}
where we have chosen $\boldsymbol{v}$ as $v_i=1/M$ for all $i$'s and defined $\hat{G}'\equiv \sum_{i=1}^M \hat{N}_i/M$~\cite{proctor2018multiparameter}, 
and $\max_{\psi}$ denotes the maximization over Gaussian states with a photon number constraint $\langle \psi|\sum_{j=1}^M\hat{N}_j|\psi\rangle=\bar{N}$. 
Here, the second inequality uses the Cauchy-Schwarz inequality, $(\boldsymbol{w}^\text{T}\bm{H}^{-1}\boldsymbol{w})(\boldsymbol{v}^\text{T}\bm{H}\boldsymbol{v})\geq(\boldsymbol{w}\cdot \boldsymbol{v})^2$, with the equality condition being $\bm{H} \boldsymbol{v}\propto \boldsymbol{w}$, which will be used to show the tightness of the inequality. In addition, we have used the fact that $(\boldsymbol{v}^\text{T}\bm{H}\boldsymbol{v})^{-1}$ is equivalent to the QCRB for estimating a parameter generated by $\hat{G}'$ \cite{proctor2018multiparameter}.
The third inequality is obtained via the maximization over Gaussian states with a given photon number $\bar{N}$.
For the final equality, we have used $\boldsymbol{w}\cdot \boldsymbol{v}=\|\boldsymbol{w}\|_1/M$ and $\max_\psi (\Delta^2 \hat{G}')_\psi=2\bar{N}(\bar{N}+1)/M^2$ \cite{oh2020optimal}. 
This result is used as the individual lower bounds of $\Delta^2 \phi_\pm^{*}$ in Eq.~\eqref{eq:err_sum}.
Note that here we do not assume the first moment vector to be zero in deriving the QCRB of Eq.~\eqref{eq:QCRB_individual}, so applicable to Gaussian states with non-zero displacement.
Furthermore, it is worth emphasizing that the optimal estimation error derived in Eq.~\eqref{eq:QCRB_individual} holds even if we include an ancillary system that does not pass through phase shifters.
Thus, the ultimate bound of Eq.~\eqref{eq:QCRB_individual} is valid for any Gaussian probe state.

It is interesting to note that the ultimate estimation error of Eq.~\eqref{eq:QCRB_individual} is the same for all the cases when $w_i$'s have the equal sign. It does not depend on the magnitude distribution $\{w_i\}_{i=1}^M$, but its norm $\| \boldsymbol{w}\|_1^2$. Furthermore, the bound of Eq.~\eqref{eq:QCRB_individual} for arbitrary $\{w_i\}_{i=1}^M$ generalizes the previous result that has been found for $w_i=1/M$~$\forall i$~\cite{oh2020optimal}.


We further emphasize that the bound of Eq.~\eqref{eq:QCRB_individual} can be achieved by a single-mode squeezed vacuum state injected into a BSN whose parameters are determined by the magnitude distribution $\{w_i\}_{i=1}^M$. One example of a BSN to implement it is written as 
\begin{align}
    \hat{U}_\text{BSN}=\hat{B}_{M-1,M}(\theta_{M-1})\hat{B}_{M-2,M-1}(\theta_{M-2})\cdots \hat{B}_{1, 2}(\theta_1)
\end{align}
where $\hat{B}_{i,j}(\theta_j)=\exp[\theta_j(\hat{a}_i^{\dagger}\hat{a}_{j}-\hat{a}_i\hat{a}_{j}^{\dagger})]$ and $\theta_j=\arccos(w_j/\|\boldsymbol{w}\|_1\prod_{k=0}^{j-1}{\sin^2 \theta_k})^{1/2}$ with defining $\theta_0=\pi/2$ 
(see Appendix \ref{appendix:opt_state} for the details). 
For the multi-mode Gaussian probe state prepared as above, the homodyne detection is the optimal measurement setting to reach the bound of Eq.~\eqref{eq:QCRB_individual} (see Appendix~\ref{appendix:homodyne} for the details). There, the homodyne angle needs to be set to $\theta_\text{HD}^i=\phi_i-\arccos(\tanh{2r})$ for a squeezing parameter $r$ of the input squeezed vacuum state. 

Therefore, using the aforementioned optimal scheme individually for the two groups in the entire scheme for arbitrary positive and negative weights, the error bound of Eq.~\eqref{eq:err_sum} can be achieved in practice. 

\subsection{Optimality for the entire scheme with arbitrary weights}\label{sec:optimal}
Now, we prove that the proposed scheme separating the two groups as shown in Fig.~\ref{proposed_setup} is indeed optimal when Gaussian states with zero displacement are used. For the purpose, consider an estimation of $\phi^*$ under the condition that $\phi_\pm^*=\boldsymbol{w}_\pm^\text{T}\boldsymbol{\phi}$ are unknown but the other $(M-2)$ parameters $\tilde{\phi}=\boldsymbol{u}^\text{T}\boldsymbol{\phi}$ are known for $\boldsymbol{u}$ being linearly independent of $\boldsymbol{w}_\pm$. Note that the estimation of $\phi^*$ when $\tilde{\phi}$'s are known is obviously easier than the case when all the other parameters are unknown. Thus, the optimal estimation error of the latter (harder), which is the focus of this work, cannot be smaller than that of the former task (easier), i.e., $\Delta\phi_{\text{easy}}^{*\text{QCRB}}\le\Delta\phi_\text{hard}^{*\text{QCRB}}$ (see its formal proof in Appendix~\ref{appendix:comparison}).
Below, we derive the optimal estimation error $\Delta\phi^{*\text{QCRB}}_\text{easy}$ of the easier case and then show that our proposed scheme can achieve it, i.e., $\Delta\phi^{*\text{QCRB}}_\text{easy}=\Delta\phi^{*\text{QCRB}}_\text{hard}$. 
We note that the choice of $\boldsymbol{u}$ does not change the analysis below since the knowledge of $\tilde{\phi}$'s for a given basis can be converted to the one in a different choice of basis \cite{paris2009quantum, proctor2018multiparameter}.

Let us derive the optimal estimation error for $\phi^*$ when $(M-2)$ parameters of $\tilde{\phi}$'s are all known, i.e., $\Delta\phi^{*\text{QCRB}}_\text{easy}$.
To do that, consider the relevant QFIM for $\phi_\pm^*$, which reads $\tilde{H}_{\alpha\beta}=2\langle\{\hat{G}_\alpha-\langle\hat{G}_\alpha \rangle,\hat{G}_\beta-\langle\hat{G}_\beta\rangle\}\rangle$ for $\alpha,\beta\in \{+,-\}$, where $\hat{G}_+=\sum_{i:w_i>0}w_i\hat{a}_i^\dagger\hat{a}_i$ and $\hat{G}_-=\sum_{i:w_i<0}w_i\hat{a}_i^\dagger\hat{a}_i$.
One can then show 
\begin{align}
    \tilde{H}_{+-}=\tilde{H}_{-+}=4\sum_{i:w_i>0}\sum_{j:w_j<0}w_iw_jC(\hat{N}_i,\hat{N}_j)\leq 0,
    \label{eq:Hpm_Negative}
\end{align}
where $w_iw_j<0$ and 
\begin{align}
    C(\hat{N}_i,\hat{N}_j)&\equiv \langle \hat{N}_i\hat{N}_j \rangle-\langle\hat{N}_i\rangle\langle\hat{N}_j\rangle \\ 
&=2\left(\langle \hat{x}_i\hat{x}_j\rangle^2+\langle \hat{x}_i\hat{p}_j\rangle^2+\langle \hat{p}_i\hat{x}_j\rangle^2+\langle \hat{p}_i\hat{p}_j\rangle^2\right)\geq 0
\end{align}
for Gaussian states with zero displacement.
Note that $C(\hat{N}_i,\hat{N}_j)$ represents the photon-number correlation function and that the equality $C=0$ holds if and only if the Gaussian state is a product state.
The QCRB for $\phi^*=\phi^*_++\phi^*_-$ can thus be written as
\begin{align}
    \Delta^2\phi_\text{easy}^*&\geq 
    [\tilde{\boldsymbol{H}}^{-1}]_{11}+[\tilde{\boldsymbol{H}}^{-1}]_{22}+2[\tilde{\boldsymbol{H}}^{-1}]_{12} \\ 
    &\geq [\tilde{\boldsymbol{H}}^{-1}]_{11}+[\tilde{\boldsymbol{H}}^{-1}]_{22} \\ 
    &\geq \frac{\|\boldsymbol{w}_+\|_1^2}{8\bar{N}_+(\bar{N}_++1)}+\frac{\|\boldsymbol{w}_-\|_1^2}{8\bar{N}_-(\bar{N}_-+1)}
    \equiv \Delta^2\phi_\text{easy}^{*\text{QCRB}}, \label{eq:easy_QCRB}
\end{align}
where we have used Eq.~\eqref{eq:Hpm_Negative} for the second inequality and Eq.~\eqref{eq:QCRB_individual} for the third inequality.
Note that the minimized QCRB for the easier task is the same as Eq.~\eqref{eq:err_sum}.
This means that our proposed scheme achieves the optimal error bound to $\Delta^2\phi_\text{easy}^*$ for estimating $\phi^*$ when $\tilde{\phi}$'s are fixed and known.
More formally, we have
\begin{align}\label{eq:chain}
    \Delta^2\phi_\text{easy}^{*\text{QCRB}}\leq
    \Delta^2\phi_\text{hard}^{*\text{QCRB}}\leq
    \frac{\|\boldsymbol{w}_+\|_1^2}{8\bar{N}_+(\bar{N}_++1)}+\frac{\|\boldsymbol{w}_-\|_1^2}{8\bar{N}_-(\bar{N}_-+1)},
\end{align}
where the first inequality is trivial, and the second inequality is from the fact that $\Delta^2 \phi_\text{hard}^{*\text{QCRB}}$ is lower than an estimation error of a particular scheme [see Eq.~\eqref{eq:err_sum}].
Interestingly, the upper bound of Eq.~\eqref{eq:chain}, which is set by the proposed scheme, is equivalent to the lower bound of Eq.~\eqref{eq:chain}. 
This means that Eq.~\eqref{eq:err_sum} is the QCRB for an arbitrary scheme using Gaussian states with zero displacement.

One may wonder what if we use a globally entangled state naively prepared considering only the weight magnitudes while ignoring their signs, i.e., the scheme shown in Fig.~\ref{setting}. In Appendix~\ref{appendix:ignored_signs}, we show that the scheme distributing a squeezed vacuum state via the BSN according to the weight magnitudes leads to the same error bound as the SQL of Eq.~\eqref{eq:SQL}, i.e., even worse than the case using optimal product Gaussian states leading to the bound of Eq.~\eqref{OPGSerror}. Therefore, the scheme distributing the squeezed vacuum into multiple modes is not useful when opposite weight signs are involved. For the latter, it is clear that the scheme in Fig.~\ref{proposed_setup} is optimal, but Heisenberg-limited suboptimal schemes can also be found as in Ref.~\cite{Malitesta2021}.

One can also notice that the optimal scheme using Gaussian states with zero displacement estimates $\phi_{+}^{*}=\boldsymbol{w}_{+}^\text{T}\boldsymbol{\phi}$ and $\phi_{-}^{*}=\boldsymbol{w}_{-}^\text{T}\boldsymbol{\phi}$ individually and then combine them via post-data processing. Such a treatment leads to no difference in the estimation uncertainty between the estimation of $\phi_{+}^{*}+\phi_{-}^{*}$ and $\phi_{+}^{*}-\phi_{-}^{*}$.
However, the estimation of $\phi_{+}^{*}-\phi_{-}^{*}$ boils down to the average phase estimation with equal weight signs, i.e., $(\boldsymbol{w}_{+}^\text{T}+\vert \boldsymbol{w}_{-}^\text{T}\vert)\boldsymbol{\phi}$. In this case, the scheme in Fig.~\ref{proposed_setup} is not optimal; the scheme distributing a squeezed vacuum state over multiple modes via the BSN is optimal as we have shown in both the present work [see Eq.~\eqref{eq:QCRB_individual}] and our previous work in Ref.~\cite{oh2020optimal}. Therefore, 
it should be noted that the scheme in Fig.~\ref{proposed_setup} is only advantageous when estimating the sum of the individual subglobal parameters, $\phi_{+}^{*}+\phi_{-}^{*}$.

\subsection{Two-mode scheme for arbitrary weights}\label{sec:two-mode}
\textit{Gaussian probe state -- }
To have a better understanding, let us concentrate here on a two-mode distributed sensing scheme using Gaussian states with zero displacement to estimate $\phi^{*}=w_1\phi_1+w_2\phi_2$ for arbitrary $w_1$ and $w_2$. In such a scheme, the QCRB of Eq.~\eqref{err} can be written as
\begin{align}
\Delta^2 \phi^{*}
&\geq w_1^2[\boldsymbol{H}^{-1}]_{11}+w_2^2[\boldsymbol{H}^{-1}]_{22}+2w_1 w_2[\boldsymbol{H}^{-1}]_{12},\label{eq:QCRB_two_mode}
\end{align}
where $[\boldsymbol{H}^{-1}]_{12}=[\boldsymbol{H}^{-1}]_{21}$ has been used. 
In this case, $H_{12}=4C(\hat{N}_1,\hat{N}_2)>0$ for all two-mode Gaussian states with zero displacement, leading to
$[\boldsymbol{H}^{-1}]_{12}\le 0$ in Eq.~\eqref{eq:QCRB_two_mode}. One can now see that the last term in Eq.~\eqref{eq:QCRB_two_mode} becomes positive (negative) when $w_1$ and $w_2$ have opposite (equal) signs. This implies that entanglement between the two modes is detrimental when the weight signs are opposite, whereas is advantageous otherwise. 

\begin{figure}[t]
\centering
\includegraphics[width=0.95\linewidth]{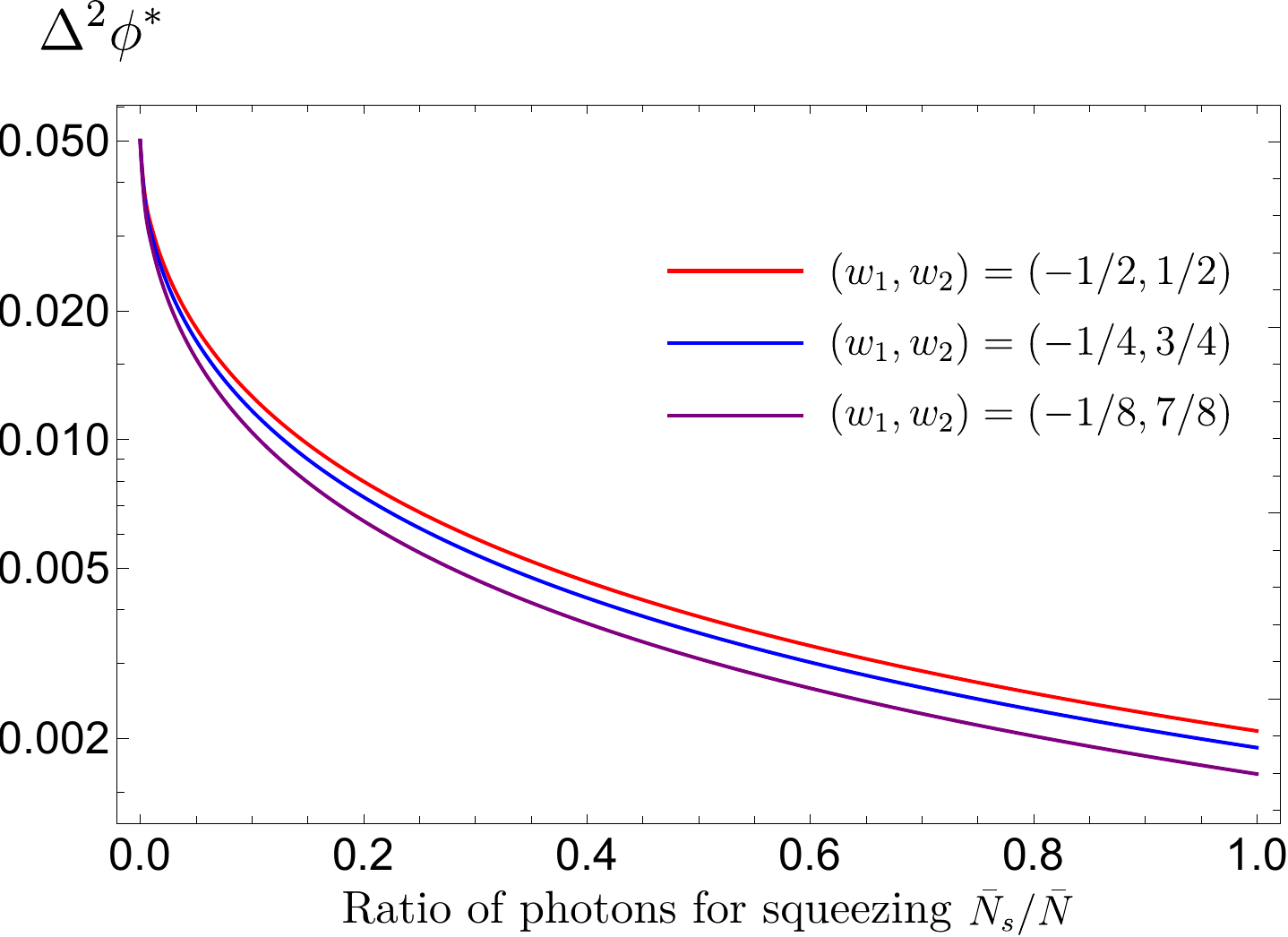}
\caption{Numerically minimized QCRB $\Delta^2 \phi^*$ for different ratios of the average energy allocated for squeezing, $\bar{N}_\text{s}$, to the total average energy, $\bar{N}$. 
As example, three cases of the weights $(w_1,w_2)$ are considered for a given $\bar{N}=10$. 
It is clear that the estimation error bound is minimized when all the energies are spent for squeezing, i.e., when $\bar{N}_\text{s}/\bar{N}=1$. Also note that in any case, the optimized BSN turns out to be an identity, implying that no entanglement is used in the optimized two-mode scheme when the weight signs are opposite.
}
\label{ratio}
\end{figure}

Instead of directly applying the above approach to Gaussian states with non-zero displacement, we alternatively perform numerical optimization over all two-mode Gaussian probe states to minimize the QCRB. Suppose that two arbitrary single-mode Gaussian pure states (i.e., squeezed displaced states) are injected into a beam splitter and then undergo the phase shifts. For a given total energy $\bar{N}$ and weights $(w_1,w_2)$, the squeezing parameters $(\xi_1=r_1 e^{i\varphi_1},\xi_2=r_2 e^{i\varphi_2})$, the displacement parameters $(\alpha_1,\alpha_2)$, and the beam splitter parameter $\theta$ are optimized while keeping a certain ratio between $\bar{N}_\text{s}=\sinh^2 r_1+\sinh^2r_2$ and $\bar{N}_\text{d}=|\alpha_1|^2+|\alpha_2|^2$ (i.e., $\bar{N}=\bar{N}_\text{s}+\bar{N}_\text{d}$).
Hence, the probe state to be optimized reads 
\begin{align}
|\psi_\text{in}\rangle=e^{i\theta(\hat{a}_{1}\hat{a}_{2}^\dagger-\hat{a}_{1}^\dagger\hat{a}_{2})}
\hat{D}_{1}(\alpha_1)\hat{S}_{1}(\xi_1)\hat{D}_{2}(\alpha_2)\hat{S}_{2}(\xi_2)|0\rangle,
\end{align}
where 
$\hat{a}_{j}$ is the annihilation operator in $j$th mode, and $\hat{S}(\cdot)$ and $\hat{D}(\cdot)$ are the squeezing and displacement operators, respectively.
Figure~\ref{ratio} presents the minimized QCRB in terms of the ratio of $\bar{N}_\text{s}$ to $\bar{N}$ for $\bar{N}=10$ and three example cases of $(w_1,w_2)$. It clearly shows that the QCRB becomes smaller as the contribution of displacement is reduced, i.e., Gaussian states with zero displacement are optimal. We also emphasize that the optimized BSN for all cases shown in Fig.~\ref{ratio} turns out to be an identity (not shown), implying that the optimized scheme does not exploit entanglement even when displacement is involved.
It is worth noting that the estimation error increases only slightly when a small portion of photons is allocated for displacement for a given $\bar{N}$ (see gradual curves around $\bar{N}_\text{s}/\bar{N} =1$ in Fig.~\ref{ratio}).

As analyzed above, for the case of the opposite signs, the best choice is to remove the correlation, leading to $[\boldsymbol{H}^{-1}]_{12}=0$ in Eq.~\eqref{eq:QCRB_two_mode}. Furthermore, the diagonal terms can be minimized, under the photon number constraint, by employing a product state of two single-mode squeezed vacua, for which the QCRB is written as
\begin{align}\label{eq:twoQCRB}
\Delta^2 \phi^{*}
\geq \frac{w_1^2}{8\bar{N}_1(\bar{N}_1+1)}+\frac{w_2^2}{8\bar{N}_2(\bar{N}_2+1)},
\end{align}
where the optimal condition for $\bar{N}_i$ in each mode is again $\bar{N}_i^2(\bar{N}_i+1)^2/(2\bar{N}_i+1)\propto w_i^2$. 
Again, as in Eq.~\eqref{eq:min}, the QCRB in Eq.~\eqref{eq:twoQCRB} is upper-bounded by $1/4\bar{N}^2$, i.e., it achieves the Heisenberg scaling in $\bar{N}$.
The optimality of the two single-mode squeezed vacua input has been similarly identified when only the phase-difference is unknown in a two-mode interferometer~\cite{lang2014optimal}. To avoid a wrong impression that Gaussian entanglement between the two modes is useless in phase-difference sensing, we stress that the Heisenberg-limited QCRB can be achieved by entangling a coherent state and a single-mode squeezed vacuum via a beam splitter~\cite{lang2013} although the product Gaussian states without displacement is optimal for a given total energy constraint.
For the case of the same signs, on the other hand, the optimal scheme is to distribute a single squeezed vacuum state to the two modes via a beam splitter, so as to manipulate the interplay between the photon-number fluctuation of each mode and the photon-number correlation between the modes. It finally leads to the QCRB of Eq.~\eqref{eq:QCRB_individual}. 
It is thus clear that the optimal two-mode phase sensing scheme depends on the signs of $w_i$'s as well as their magnitudes.

\textit{Non-Gaussian probe state -- }
Note, however, that the above behavior may change when non-Gaussian probe states are used. 
Let us here consider particularly a quantum state with the maximum photon number of $N$ as an example for estimation of the global parameter $\phi_\pm=(\phi_1 \pm \phi_2)/2$.
Note that the constraint of the maximum photon number $N$ is further imposed to non-Gaussian probe states for conciseness in addition to the total average photon number constraint $\bar{N}$ being considered throughout this work.
For the estimation of $\phi_-$, one can find that the optimal entangled state in the absence of loss is a so-called NOON state, which reads
\begin{align}
    |\psi_\text{NOON}\rangle=\frac{1}{\sqrt{2}}(|N0\rangle+|0N\rangle),
\end{align}
for which $\bar{N}_{1,2}=N/2$, such that $\bar{N}=N$.
For this state, the correlation function reads (see Appendix \ref{appendix:noon} for the detail)
\begin{align}
C_\text{NOON}=-\frac{\bar{N}^2}{4}.
\end{align}
The latter clearly shows that the crucial feature to enhance the sensitivity in the estimation of $\phi_-$ is the anti-correlation of photon number between the two modes. When estimating $\phi_+$, on the other hand, one can similarly show that the best strategy is to employ a photon-number-correlated state written as
\begin{align}
    |\psi_\text{NNOO}\rangle=\frac{1}{\sqrt{2}}(|NN\rangle+|00\rangle),
\end{align}
for which
\begin{align}
C_\text{NNOO}=\frac{\bar{N}^2}{4}.
\end{align}
The above example shows that advantageous (non-Gaussian) entangled states can be found for arbitrary positive and negative weights. 
It stems from the fact that photon-number correlated or anti-correlated non-Gaussian states are all available. 
Such a property, however, does not exist in the class of zero-mean Gaussian states, for which all the off-diagonal elements of QFIM are non-negative. 
The latter feature thus explains why reducing the quantum correlation between Gaussian states is more advantageous than enhancing it in estimating a linearly combined parameter with opposite weight signs~\cite{lang2014optimal} although the latter scheme is still helpful~\cite{lang2013}.



\section{Discussion}
We have proposed a Gaussian quantum phase sensing scheme using Gaussian probe states with no displacement for a global parameter defined as a linear combination of multiple phases with arbitrary positive and negative weights.
We have shown that the scheme is optimal among zero-mean Gaussian probe states.
The scheme divides the modes into two groups based on the sign of corresponding weights, and uses entangled input resources only within the individual groups, i.e., no entanglement between the two separate groups. 
Such an interesting feature has been understood by elaborating on the two-mode distributed sensing scenario and comparing it with the cases using entangled photons. Particularly for the two-mode case, we have numerically demonstrated that the optimal Gaussian scheme exploits neither entanglement between the two groups nor displacement at all. 
To be accurate, Gaussian entanglement is still helpful~\cite{lang2013} for outperforming classical schemes, but it is better to reduce its strength from the perspective of optimality for a given total energy constraint.

It is worth noting that another scheme of estimating arbitrary linear combination of phases has been proposed to achieve the Heisenberg scaling using a Gaussian state input \cite{Malitesta2021}.
Their scheme is to distribute a squeezed vacuum state over $M$ modes using a BSN, and the $M$-mode Gaussian output state is fed into the respective Mach-Zehnder interferometers with an additional coherent state for each mode followed by photon-number detection.
A crucial difference from our scheme is that their scheme uses an entanglement over the $M$ modes regardless of the signs of weights.
Interestingly, it achieves the Heisenberg scaling using Gaussian entanglement over the $M$ modes regardless of the signs of the weights.
It implies that while our scheme is optimal for zero-mean Gaussian states, there still exist other schemes that attain the Heisenberg scaling.

It is interesting to further study the effect of loss in the optimal error bound of the proposed scheme. 
The most interesting question would be whether or not the scheme we propose in this work is ultimately optimal even for Gaussian states with non-zero displacement. 
We think that it is likely to be the case as in single-parameter estimation, where the optimal scheme using Gaussian states does not employ displacement for a given total energy~\cite{matsubara2019optimal}.
We however leave its proof as future study due to the complexity of the analysis required.
Whether the Heisenberg scaling is maintained with reducing the ratio of photons for squeezing in the optimal scheme needs to be elaborated in future study in that displacing is easier than squeezing from a practical perspective.
Similar questions addressed in this work can also be asked for other kinds of physical parameters such as displacement or intensity. Moreover, the use of non-Gaussian probe states can be considered and compared with schemes using only Gaussian states. 

\section*{acknowledgments}
C.O. and L.J. acknowledge support from the ARO (W911NF-18-1-0020, W911NF-18-1-0212), ARO MURI (W911NF-16-1-0349), AFOSR MURI (FA9550-19-1-0399, FA9550-21-1-0209), DoE Q-NEXT, NSF (EFMA-1640959, OMA-1936118, EEC-1941583), NTT Research, and the Packard Foundation (2013-39273).
C.L. is supported by a KIAS Individual Grant (QP081101) via the Quantum Universe Center at Korea Institute for Advanced Study and Korea Research Institute of Standards and Science (KRISS–GP2022-0012).

\appendix
\begin{widetext}
\section*{Appendix}
\section{Upper bound of the Quantum Cram\'{e}r-Rao bound of two squeezed vacuum scheme}\label{appendix:upper}
In this Appendix, we show that the quantum Cram\'{e}r-Rao bound (QCRB) of our scheme is upper-bounded by the Heisenberg scaling in $\bar{N}$.
Recall the QCRB in Eq.~\eqref{eq:err_sum}. It can be generally written as 
\begin{align}
    \frac{\|\boldsymbol{w}_+\|_1^2}{8\bar{N}_+(\bar{N}_++1)}+\frac{\|\boldsymbol{w}_-\|_1^2}{8\bar{N}_-(\bar{N}_-+1)}< \frac{\|\boldsymbol{w}_+\|_1^2}{8\bar{N}_+^{2}}+\frac{\|\boldsymbol{w}_-\|_1^2}{8\bar{N}_-^{2}}
\end{align}
for arbitrary $\bar{N}_\pm$ such that $\bar{N}_+ +\bar{N}_- =\bar{N}$. Optimizing $\bar{N}_\pm$ for each expression, one can minimize each side individually, so that 
\begin{align}\label{eq:minimize}
    \min_{\{\bar{N}_\pm\}}\left(\frac{\|\boldsymbol{w}_+\|_1^2}{8\bar{N}_+(\bar{N}_++1)}+\frac{\|\boldsymbol{w}_-\|_1^2}{8\bar{N}_-(\bar{N}_-+1)}
    \right)
    < 
    \min_{\{\bar{N}_\pm\}}\left(\frac{\|\boldsymbol{w}_+\|_1^2}{8\bar{N}_+^{2}}+\frac{\|\boldsymbol{w}_-\|_1^2}{8\bar{N}_-^{2}}\right).
\end{align}
The minimization can be done via the Lagrange multiplier method under the constraint $\bar{N}_++\bar{N}_-=\bar{N}$, consequently leading to the optimal energy allocation for each: $\bar{N}_\pm^{*}$ satisfying 
$\bar{N}_\pm^{*2}(\bar{N}_\pm^*+1)^2/(2\bar{N}_\pm^*+1)\propto \|\boldsymbol{w}_\pm\|_1^2$ for the left-hand-side and $\bar{N}_\pm=\bar{N}\|\boldsymbol{w}_\pm\|_1^{2/3}/(\|\boldsymbol{w}_+\|_1^{2/3}+\|\boldsymbol{w}_-\|_1^{2/3}) $ for the right-hand-side. Plugging the latter solution to Eq.~\eqref{eq:minimize}, one can find the upper bound of the minimized QCRB written as
\begin{align}
    \frac{\|\boldsymbol{w}_+\|_1^2}{8\bar{N}_+^*(\bar{N}_+^*+1)}+\frac{\|\boldsymbol{w}_-\|_1^2}{8\bar{N}_-^*(\bar{N}_-^*+1)}
    < 
    \min_{\{\bar{N}_\pm\}}\left(\frac{\|\boldsymbol{w}_+\|_1^2}{8\bar{N}_+^{2}}+\frac{\|\boldsymbol{w}_-\|_1^2}{8\bar{N}_-^{2}}\right)
    = \frac{\|(\|\boldsymbol{w}_+\|_1,\|\boldsymbol{w}_-\|_1)\|^{2}_{2/3}}{8\bar{N}^2}\leq \frac{1}{4\bar{N}^2},
\end{align}
where we have used the inequality between $p$-norms, $\|\bm{x}\|_p\leq n^{1/p-1/q}\|\bm{x}\|_q$ for $0<p\leq q<\infty$ and $\|\bm{x}\|_p\equiv (\sum_i x_i^p)^{1/p}$ for a $n$-dimensional vector $\bm{x}$. 
Hence, the minimized QCRB is upper-bounded as
\begin{align}
    \frac{\|\boldsymbol{w}_+\|_1^2}{8\bar{N}_+^*(\bar{N}_+^*+1)}+\frac{\|\boldsymbol{w}_-\|_1^2}{8\bar{N}_-^*(\bar{N}_-^*+1)}
    < \frac{1}{4\bar{N}^2}.
\end{align}

\section{Optimal Gaussian state for arbitrary weights with an equal sign}\label{appendix:opt_state}
Here, we find the optimal state to estimate a linear combination of phases for arbitrary weights with an equal sign, namely, the elements of the weight vector $\boldsymbol{w}$ are all positive. Let us begin with rewriting the elements of quantum Fisher information matrix (QFIM) of Eq.~\eqref{qfim} for Gaussian states without displacement as
\begin{align}
H_{ij}&=2\text{Tr}[P_iO\bm{\Gamma}_\text{in}O^\text{T}P_j P_jO\bm{\Gamma}_\text{in}O^\text{T}P_i]-\delta_{ij}
=2\text{Tr}[O^\text{T}P_iO\bm{\Gamma}_\text{in}O^\text{T}P_j O\bm{\Gamma}_\text{in}]-\delta_{ij},
\end{align}
where $P_i\equiv |i\rangle\langle i|\otimes \mathbb{1}_2$ is a projector,
$O\equiv \tilde{O}\otimes \mathbb{1}_2$ is the first beam splitter network (BSN), and $\bm{\Gamma}_\text{in}$ is the covariance matrix of an input state.
The covariance matrix of an input state assumed to be a product state of a squeezed vacuum state and $(M-1)$ vacua can be written as
\begin{align}
\bm{\Gamma}_\text{in}&=|1\rangle\langle 1| \otimes D +\sum_{n=2}^M |n\rangle\langle n| \otimes \frac{\mathbb{1}_2}{2},
\end{align}
where $D=\frac{1}{2}\text{diag}(e^{2r},e^{-2r})$. 

First, let us show that the above state satisfies $\bm{H}\boldsymbol{v}\propto \boldsymbol{w}$ with $v_i=1/M$, corresponding to the equality condition of the second Cauchy-Schwarz inequality in Eq.~\eqref{eq:QCRB_individual}.
When $v_i=1/M$, the vector $\bm{H}\boldsymbol{v}$ can be developed as follows. 
\begin{align}
(\bm{H}\boldsymbol{v})_i
&=\frac{1}{M}\sum_{j=1}^M H_{ij}\\
&\propto \sum_{j=1}^M \left(2\text{Tr}[O^\text{T}P_iO\bm{\Gamma}_\text{in}O^\text{T}P_j O\bm{\Gamma}_\text{in}]-\delta_{ij}\right) \\
&=2\text{Tr}[O^\text{T}P_iO\bm{\Gamma}_\text{in}^2]-1 \\
&=2\text{Tr}\left [\left(\tilde{O}^\text{T}|i\rangle\langle i|\tilde{O}\otimes \mathbb{1}_2\right) \left(|1\rangle\langle 1|\otimes D^2+\sum_{n=2}^M |n\rangle\langle n|\otimes \frac{1}{4}\mathbb{1}_2\right)\right]-1 \\
&=2\text{Tr}[D^2]\langle1|\tilde{O}^\text{T}|i\rangle\langle i|\tilde{O}|1\rangle+\sum_{n=2}^M \langle n|\tilde{O}^\text{T}|i\rangle\langle i|\tilde{O} |n\rangle -1 \\
&=2\text{Tr}[D^2]\langle1|\tilde{O}^\text{T}|i\rangle\langle i|\tilde{O}|1\rangle+\langle i|\tilde{O} (\mathbb{1}_M-|1\rangle\langle1|)\tilde{O}^\text{T}|i\rangle -1 \\
&=(2\text{Tr}[D^2]-1)\langle1|\tilde{O}^\text{T}|i\rangle\langle i|\tilde{O}|1\rangle\\
&=(2\text{Tr}[D^2]-1)|\langle i|\tilde{O}|1\rangle|^2.
\end{align}
It is thus clear that $\bm{H}\boldsymbol{v}\propto \boldsymbol{w}$ if the first BSN operator $O$ is constituted such that $|\langle i| \tilde{O}|1\rangle|^2=w_i/\|\boldsymbol{w}\|_1$. 
More specifically, a particular example BSN setup to satisfy the above condition can be written as $\hat{U}_\text{BSN}=\hat{B}_{M-1,M}(\theta_{M-1})\hat{B}_{M-2,M-1}(\theta_{M-2})\times \cdot\cdot\cdot \times \hat{B}_{1, 2}(\theta_1)$,
where $\hat{B}_{i,j}(\theta_j)=\exp[\theta_j(\hat{a}_i^{\dagger}\hat{a}_{j}-\hat{a}_i\hat{a}_{j}^{\dagger})]$ and $\theta_j=\arccos\left[w_j/\left(\|\boldsymbol{w}\|_1\prod_{k=0}^{j-1}{\sin^2 \theta_k}\right)\right]^{1/2}$ with defining $\theta_0=\pi/2$. 

Second, let us show that $\Delta^2\hat{G}'$ can be maximized by the probe state prepared in the aforementioned setup, satisfying the equality condition of the third inequality in Eq.~\eqref{eq:QCRB_individual}. 
Recall that $\hat{G}'=\sum_{i=1}^M v_i\hat{N}_i=\sum_{i=1}^M \hat{N}_i/M$ when $v_i=1/M$. 
In addition, it can be easily shown that the sum of photon number operators $\sum_{i=1}^M\hat{N}_i$ is invariant under the BSN operation and that the maximum photon number variance is then attained by a product of the single-mode squeezed vacuum state and $(M-1)$ vacua.
Thus, the proposed probe state maximizes $\Delta^2\hat{G}'$.

\section{Comparison of QCRBs with and without additional information} \label{appendix:comparison}
Here, we show the QCRB for $\phi^*$ when $(M-2)$ parameters $\tilde{\phi}$'s are all known except $\phi_\pm^*$ is upper-bounded by that for $\phi^*$ when all parameters are unknown.
Intuitively, this is obvious because the former has more information than the latter, so that it is easier than the latter, i.e., $\Delta\phi^*_\text{easy}\le\Delta\phi^*_\text{hard}$.
To explicitly show it, consider an $M\times M$ weight matrix $\bm{W}$ that consists of linearly independent weight vectors: $\boldsymbol{w}_\pm$ and $\boldsymbol{u}$. It maps $M$ parameters $\{\phi_i\}$ into $M$ global parameters, i.e., $\phi_\pm^*=\boldsymbol{w}_\pm^\text{T}\boldsymbol{\phi}$ and $\tilde{\phi}=\boldsymbol{u}^\text{T}\boldsymbol{\phi}$. For $M$ global parameters, the QFIM can be partitioned into four block matrices as
\begin{align}
\boldsymbol{H}=
\begin{pmatrix}
\boldsymbol{H}^\text{(A)} & \boldsymbol{H}^\text{(AB)} \\ 
\boldsymbol{H}^\text{(BA)} & \boldsymbol{H}^\text{(B)}
\end{pmatrix},
\end{align}
where the super-indices $\text{A}$ and $\text{B}$ denote the first two dimensions and the rest $(M-2)$ dimensions, respectively. When the $(M-2)$ parameters $\tilde{\phi}$'s are all known, $\boldsymbol{H}^\text{(A)}$ is the QFIM for $\phi_\pm^*$. Therefore, to show $\Delta\phi^*_\text{easy}\le\Delta\phi^*_\text{hard}$ is equivalent to proving $[\boldsymbol{H}^{-1}]^\text{(A)}\geq [\boldsymbol{H}^\text{(A)}]^{-1}$, and it can be verified by applying an analytical blockwise inversion formula to $\boldsymbol{H}^{-1}$:
\begin{align}
[\boldsymbol{H}^{-1}]^\text{(A)}=[\boldsymbol{H}^\text{(A)}-\boldsymbol{H}^\text{(AB)}(\boldsymbol{H}^\text{(B)})^{-1}\boldsymbol{H}^\text{(BA)}]^{-1}\geq [\boldsymbol{H}^\text{(A)}]^{-1},
\end{align}
where the inequality comes from $\boldsymbol{H}^\text{(AB)}(\boldsymbol{H}^\text{(B)})^{-1}\boldsymbol{H}^\text{(BA)}\geq 0$.

\section{Globally entangled Gaussian state ignoring the weight signs}\label{appendix:ignored_signs}
Let us consider the case where $w_i=1/M$ for $1\leq i\leq M/2$ and $w_i=-1/M$ for $M/2<i\leq M$, assuming $M$ to be even for this example. Ignoring the weight signs, one can employ the optimal scheme that has previously been found for $w_i=1/M$~$\forall i$~\cite{oh2020optimal}, which uses a single-mode squeezed vacuum state input into a balanced BSN. It can be shown that the estimation error bound is given as
\begin{align}
    \Delta^2\phi^*\geq \frac{1}{4\bar{N}}=\frac{1}{4M\bar{n}},
\end{align}
where we have introduced a parameter $\bar{n}\equiv \bar{N}/M$, representing the photon number allocated in each mode.
Notably, the error bound is the same as the SQL, and even worse than the case using product Gaussian states, which is written from Eq.~\eqref{OPGSerror} as
\begin{align}
    \Delta^2\phi^*\geq \frac{M}{8\bar{N}(\bar{N}+M)}=\frac{1}{8M\bar{n}(\bar{n}+1)}.
\end{align}
On the other hand, the proposed scheme that uses the respective single-mode squeezed vacuum states for the individual groups reaches the error bound written as
\begin{align}
    \Delta^2\phi^*\geq \frac{1}{4\bar{N}(2\bar{N}+1)}=\frac{1}{8M\bar{n}(M\bar{n}+1)}.
\end{align}
This clearly shows the Heisenberg scaling. 
Therefore, when the opposite weight signs are involved, the previous scheme using globally entangled Gaussian states fails to gain a quantum advantage and is even worse than that using the product non-entangled Gaussian states, whereas the proposed scheme in this work achieves a quantum enhancement in comparison with the error bounds of Eqs.~\eqref{eq:SQL} and \eqref{OPGSerror}.
Such an enhancement is clear from the Heisenberg scaling with~$M$ for a fixed~$\bar{n}$.


\section{Optimality of homodyne detection}\label{appendix:homodyne}
Here, we show that homodyne detection is the optimal measurement setting when estimating a global parameter for arbitrary weights with an equal sign using two independent squeezed input states as proposed in the main text.
For simplicity, we assume that the weight vector $\boldsymbol{w}$ is normalized as $\|\boldsymbol{w}\|_1=1$.
The optimality can be demonstrated by showing that the classical Cram\'{e}r-Rao bound (CCRB) for homodyne detection is the same as the QCRB that is obtainable by the optimal measurement setting. 
To this end, we first derive the classical Fisher information matrix (CFIM) for a probability distribution of the homodyne detection outcomes and then use it to find the CCRB. 


The covariance matrix of the Gaussian probe state $\Gamma_\text{probe}$ \textit{before} phase shifters  written as (see Appendix \ref{appendix:opt_state})
\begin{align}
    \bm{\Gamma}_\text{probe}^{(i,j)}
    &=\sum_{k=1}^M \langle i|\tilde{O}|k\rangle\langle k|\tilde{O}^\text{T}|j\rangle D_k \\
    &=\langle i|\tilde{O}|1\rangle\langle 1|\tilde{O}^\text{T}|j\rangle D_1+\sum_{k=2}^M \langle i|\tilde{O}|k\rangle\langle k|\tilde{O}^\text{T}|j\rangle \frac{\mathbb{1}_2}{2} \\ 
    &=\langle i|\tilde{O}|1\rangle\langle 1|\tilde{O}^\text{T}|j\rangle D_1+\langle i|\tilde{O}(\mathbb{1}_M-|1\rangle\langle1|)\tilde{O}^\text{T}|j\rangle \frac{\mathbb{1}_2}{2}\\
    &=\sqrt{w_iw_j} \left(D_1-\frac{\mathbb{1}_2}{2}\right)+\delta_{ij}\frac{\mathbb{1}_2}{2}
\end{align}
is transformed \textit{after} phase shifters $\otimes_{i=1}^{M}\hat{R}(\phi_i)$ as
\begin{align}
    \bm{\Gamma}_\text{out}^{(i,j)}
    =R(\phi_i)\bm{\Gamma}_\text{probe}^{(i,j)}R^\text{T}(\phi_j)
    =\sqrt{w_iw_j}R(\phi_i)D_1R^\text{T}(\phi_j)+(\delta_{ij}-\sqrt{w_iw_j})\frac{R(\phi_i)R^\text{T}(\phi_j)}{2},
\end{align}
where
\begin{align}
    R(\phi)=
    \begin{pmatrix}
        \cos \phi & \sin \phi \\
        -\sin \phi & \cos \phi
    \end{pmatrix},
\end{align}
represents the symplectic transformation corresponding to a phase shifter $\hat{R}(\phi)$.
Noting that
\begin{align}
    \left[R(\phi_i)\text{diag}(d_1,d_2)R^\text{T}(\phi_j)\right]_{11}=d_1\cos\phi_i\cos\phi_j+d_2\sin\phi_i\sin\phi_j,
\end{align}
one can find the elements of the covariance matrix $\bm{\Gamma}_\text{HD}$ obtainable from homodyne detection performed \textit{along} $x$-axis, written as
\begin{align}\label{eq:CovHDelement}
    \langle i|\bm{\Gamma}_\text{HD}|j\rangle
    =\bm{\Gamma}_\text{out}^{(2i-1,2j-1)}
    =\sqrt{w_iw_j}\left(d_1\cos\phi_i\cos\phi_j+d_2\sin\phi_i\sin\phi_j\right)+(\delta_{ij}-\sqrt{w_i w_j})\frac{\cos(\phi_i-\phi_j)}{2}.
\end{align}
Its derivative with respect to $\phi_k$ can be written as
\begin{align}
    \partial_k\langle i|\bm{\Gamma}_\text{HD}|j\rangle&=\sqrt{w_iw_j}\left[\delta_{ik}\left(-d_1\sin\phi_i\cos\phi_j+d_2\cos\phi_i\sin\phi_j\right)+\delta_{jk}\left(-d_1\cos\phi_i\sin\phi_j+d_2\sin\phi_i\cos\phi_j\right)\right]\nonumber\\
    &\quad+(\delta_{ij}-\sqrt{w_iw_j})(\delta_{jk}-\delta_{ik})\frac{\sin(\phi_i-\phi_j)}{2}.\label{eq:DerivativeCov_0}
\end{align}
Here, $d_1=e^{-2r}/2$ and $d_2=e^{2r}/2$ are given from the input squeezed vacuum state with a squeezing parameter $r$. 
Note that the homodyne angle in homodyne detection is tunable and adds an additional phase to $\phi_i$, so we can treat them together by an overall phase $\phi_i$ without loss of generality. Assuming the homodyne angles are optimally chosen for the given phases such that $2\phi_i=\arccos(\tanh{2r})$ for all $i$'s, we can set $\phi_i=\phi~\forall i$ for convenience. Such an optimal angle condition further simplifies Eq.~\eqref{eq:DerivativeCov_0} as
\begin{align}
    \partial_k\langle i|\bm{\Gamma}_\text{HD}|j\rangle&=\sqrt{w_iw_j}(d_2-d_1)(\delta_{ik}+\delta_{jk})\cos\phi\sin\phi
    =\sqrt{w_iw_j}(\delta_{ik}+\delta_{jk})\frac{\tanh{2r}}{2}.
\end{align}
Thus, we have  
\begin{align}\label{eq:DerivativeCov}
    \partial_k\bm{\Gamma}_\text{HD}
    &=\frac{\tanh{2r}}{2}\sum_i\sqrt{w_iw_k}(|i\rangle\langle k|+|k\rangle\langle i|)  
    =\frac{\sqrt{w_k}\tanh{2r}}{2}(|\sqrt{w}\rangle\langle k|+|k\rangle\langle \sqrt{w}|),
\end{align}
where we have defined $|\sqrt{w}\rangle\equiv \sum_{i=1}^M \sqrt{w_i}|i\rangle$.

The inverse matrix of the covariance matrix can be obtained similarly by setting $\phi_i=\phi~\forall i$ such that $2\phi_i=\arccos(\tanh{2r})$. Equation~\eqref{eq:CovHDelement} now reads 
\begin{align}
    \langle i|\bm{\Gamma}_\text{HD}|j\rangle=\sqrt{w_iw_j}\left(d_1\cos^2\phi+d_2\sin^2\phi-\frac{1}{2}\right)+\frac{\delta_{ij}}{2}
    =\frac{1}{2}\sqrt{w_iw_j}\left(A-1\right)+\delta_{ij},
\end{align}
where $A\equiv \text{sech}{2r}$.
Thus, the covariance matrix of the resultant probability distribution obtained by homodyne detection with the optimal angles can be simply written as
\begin{align}
    \bm{\Gamma}_\text{HD}=\frac{1}{2}\left[\left(A-1\right)|\sqrt{w}\rangle\langle \sqrt{w}|+\mathbb{1}_M\right],
\end{align}
and its inverse matrix is simplified as
\begin{align}\label{eq:InverseCov}
    \bm{\Gamma}_\text{HD}^{-1}&=2\left[A^{-1}|\sqrt{w}\rangle\langle \sqrt{w}|+(\mathbb{1}_M-|\sqrt{w}\rangle\langle\sqrt{w}|)\right]
    =2\left[(A^{-1}-1)|\sqrt{w}\rangle\langle \sqrt{w}|+\mathbb{1}_M\right]
    =2\left(2\sinh^2{r}|\sqrt{w}\rangle\langle \sqrt{w}|+\mathbb{1}_M\right).
\end{align}



Substituting Eqs.~\eqref{eq:DerivativeCov} and \eqref{eq:InverseCov} into the CFIM written as
\begin{align}
    F_{ij}
    =\frac{1}{2}\text{Tr}[\bm{\Gamma}_\text{HD}^{-1}(\partial_{\phi_i}\bm{\Gamma}_\text{HD})\bm{\Gamma}_\text{HD}^{-1}(\partial_{\phi_j}\bm{\Gamma}_\text{HD})],
\end{align}
we can further develop the CFIM as
\begin{align}
    F_{ij}
    &=2[4\sinh^4r\langle \sqrt{w}|\partial_{\phi_i}\bm{\Gamma}_\text{HD}|\sqrt{w}\rangle\langle \sqrt{w}|\partial_{\phi_j}\bm{\Gamma}_\text{HD}|\sqrt{w}\rangle+4\sinh^2r\langle\sqrt{w}|\partial_{\phi_i}\bm{\Gamma}_\text{HD}\partial_{\phi_j}\bm{\Gamma}_\text{HD}|\sqrt{w}\rangle
    +\text{Tr}[\partial_{\phi_i}\bm{\Gamma}_\text{HD}\partial_{\phi_j}\bm{\Gamma}_\text{HD}]] \\ 
    &=\alpha w_iw_j+\beta w_i\delta_{ij},
\end{align}
where $\alpha\equiv\tanh^2 2r(8\sinh^4r+6\sinh^2r+1)$ and $\beta\equiv \tanh^22r\cosh2r$. 
The CFIM is thus of the form
\begin{align}
\bm{F}=\alpha |w\rangle\langle w|+\bm{B},
\end{align}
where $|w\rangle\equiv\sum_{i=1}^M w_i|i\rangle$ and $\bm{B}\equiv\beta \sum_{i,j=1}^{M}w_i\delta_{ij}|i\rangle\langle j|$.
Applying the Sherman-Morrison formula to the matrix $\bm{F}$ \cite{sherman1950adjustment, press2007numerical}, we have
\begin{align}
    \bm{F}^{-1}=(\alpha |w\rangle\langle w|+\bm{B})^{-1}=\bm{B}^{-1}-\frac{\alpha \bm{B}^{-1}|w\rangle\langle w|\bm{B}^{-1}}{1+\alpha \langle w|\bm{B}^{-1}|w\rangle}.
\end{align}
The CCRB for $\Delta^2\phi^*$ is then written as
\begin{align}
    \boldsymbol{w}^\text{T}\bm{F}^{-1}\boldsymbol{w}=\langle w|\bm{B}^{-1}|w\rangle-\alpha\frac{\langle w|\bm{B}^{-1}|w\rangle^2}{1+\alpha \langle w|\bm{B}^{-1}|w\rangle}
    =\frac{1}{\beta}-\frac{\alpha}{\beta}\frac{1}{\alpha+\beta}=\frac{1}{8\bar{N}(\bar{N}+1)},
\end{align}
where $\bar{N}=\sinh^2r$.
One can easily check that if we lift the normalization condition $\|\boldsymbol{w}\|_1=1$, the CCRB becomes
\begin{align}
    \boldsymbol{w}^\text{T}\bm{F}^{-1}\boldsymbol{w}=\frac{\|\boldsymbol{w}\|_1^2}{8\bar{N}(\bar{N}+1)}
\end{align}

As a result, the CCRB for homodyne detection is shown to be equal to the QCRB of Eq.~\eqref{eq:QCRB_individual}, implying that homodyne detection is optimal for estimation of a global parameter $\phi^*$ for arbitrary weights with an equal sign.

\section{Optimality of NOON state and NNOO state}\label{appendix:noon}
Here, we show that the NNOO state and NOON state are the optimal states achieving the maximum sensitivity to estimate $\phi_\pm=(\phi_1\pm\phi_2)/2$, respectively, when the maximum photon number is bounded to $N$.
The maximum photon number constraint allows the system to be treated as an $(N+1)$-dimensional discrete variable system.
In this case, it is well-known that the optimal state to estimate $\phi_\pm$ when its complementary parameter ($\phi_{\mp}$) is known, i.e., in single-parameter estimation, is GHZ-type states, namely, the NNOO state and NOON state, respectively~\cite{boixo2007generalized}.
One can easily show that the QFIs for single-parameter estimation of $\phi_\pm$ with the NNOO state and NOON state are given by
\begin{align}
    H_\text{NNOO}^{(S)}(\phi_+)=N^2, ~~~~ H_\text{NOON}^{(S)}(\phi_+)=N^2 \label{eq:noon},
\end{align}
respectively.
Note that the total average photon number $\bar{N}$ of the NNOO state and NOON state is equal to $N$.


On the other hand, the multiparameter estimation approach considered in this work derives the sensitivity bound for $\phi_\pm=(\phi_1\pm\phi_2)/2$ written as
\begin{align}
    \Delta^2\phi_\pm\geq\boldsymbol{w}_\pm^\text{T}\bm{H}^{-1}\boldsymbol{w}_\pm,
\end{align}
where $\boldsymbol{w}_\pm=(1,\pm 1)/2$ and the QFIM $\bm{H}$ reads
\begin{align}
    \bm{H}=
    \begin{pmatrix}
        H_{11} & H_{12} \\
        H_{21} & H_{22}
    \end{pmatrix}
\end{align}
with $H_{ij}=4(\langle \hat{N}_i\hat{N}_j \rangle-\langle\hat{N}_i\rangle\langle\hat{N}_j\rangle)$. For estimation of $\phi_+$ with the NNOO state, the QFIM elements are thus given by
\begin{align}
    H_{11}=H_{12}=H_{22}=\bar{N}^2,
\end{align}
while for estimation of $\phi_-$ with the NOON state, the QFIM elements are given by
\begin{align}
    H_{11}=H_{22}=\bar{N}^2, ~~~ H_{12}=-\bar{N}^2.
\end{align}
Notice that the QFIMs are singular in both cases. 
Thus, we project the matrices on the subspaces spanned by $\boldsymbol{w}_\pm$, respectively, resulting in
\begin{align}
    \Delta^2\phi_\pm\geq \frac{1}{\bar{N}^2}. \label{eq:multi_noon}
\end{align}

Noting that the multiparameter error bound for estimating $\phi_\pm$ is always greater than or equal to the single-parameter bound implied by Eqs.~\eqref{eq:noon} and that the above bounds \eqref{eq:multi_noon} obtained from multiparameter estimation theory are the same as Eqs.~\eqref{eq:noon} obtained from a single-parameter estimation point of view, the NNOO state and NOON state are optimal in achieving the maximum sensitivity for estimation of $\phi_\pm=(\phi_1\pm\phi_2)/2$ as well.

\end{widetext}


\bibliography{reference.bib}

\end{document}